# Semiconductor double quantum dot micromaser

Y.-Y. Liu[1], J. Stehlik[1], C. Eichler[1], M. J. Gullans[2], J. M. Taylor[2], J. R. Petta[1,3*]

[1]Department of Physics, Princeton University, Princeton, New Jersey 08544, USA.

[2]Joint Quantum Institute, University of Maryland/National Institute of Standards and Technology, College Park, Maryland 20742, USA.

[3]Department of Physics, University of California, Santa Barbara, California 93106, USA.

*Correspondence to: petta@princeton.edu

**Abstract:** The coherent generation of light, from masers to lasers, relies upon the specific structure of the individual emitters that lead to gain. Devices operating as lasers in the few-emitter limit provide opportunities for understanding quantum coherent phenomena, from THz sources to quantum communication. Here we demonstrate a maser that is driven by single electron tunneling events. Semiconductor double quantum dots (DQDs) serve as a gain medium and are placed inside of a high quality factor microwave cavity. We verify maser action by comparing the statistics of the emitted microwave field above and below the maser threshold.

A conventional laser uses an ensemble of atoms that are pumped into the excited state to achieve population inversion (*1, 2*). Enabled by advances in semiconductor device technology, semiconductor lasers quickly evolved from p-i-n junctions (*3, 4*), to quantum well structures (*5*), and quantum cascade lasers (QCLs) (*6*). In QCLs, an electrical bias is applied across exquisitely engineered multiple quantum well structures, resulting in cascaded intra-band transitions between confined two-dimensional electronic states that lead to photon emission (*7*). However, QCL emission frequencies are set by heterostructure growth profiles and cannot be easily tuned in-situ. At the same time, in atomic physics, researchers demonstrated a single atom maser, where atoms prepared in the excited state transit through a microwave cavity for a precisely controlled period of time, such that the atom "swaps" its excitation to the microwave cavity, generating a large photon field (*8*). These early experiments were extended to a single atom trapped in a high finesse optical cavity (*9*) as well as condensed matter systems, where artificial atoms were strongly coupled to cavities (*10-14*).

Here we demonstrate a maser that is driven by single electron tunneling events. The gain medium consists of semiconductor double quantum dots (DQD) that support zero-dimensional electronic states (*15*). Electronic tunneling through the DQDs generates photons that are coupled to a cavity mode (*16*). In contrast to optically pumped systems, population inversion is generated in the DQD system through the application of a bias voltage that results in sequential single electron tunneling.

The maser consists of two semiconductor DQDs (referred to as the left DQD and right DQD, as shown in Fig. 1), which are electric-dipole coupled to a microwave cavity. The cavity is formed from a half-wavelength ($\lambda/2$) superconducting Nb transmission line resonator with a center frequency $f_c = 7880.55$ MHz and a loaded quality factor $Q_c \approx 3000$ (*17, 18*). Two lithographically defined InAs nanowire DQDs serve as the maser gain medium (*16, 19*). Each DQD is fabricated by placing a single InAs nanowire over five Ti/Au bottom gate electrodes as shown in Fig. 1C (*20, 21*). The bottom gates create a tunable DQD confinement potential in the nanowire (*21*). Electrostatically defined DQDs, often regarded as artificial molecules (*15*), are a unique gain medium. They are fully reconfigurable, with electronic transitions that can be tuned from GHz to THz frequencies.

A source-drain bias voltage $V_{SD} = 2$ mV is applied across the DQDs in order to drive a current. The energy levels of each DQD can be separately tuned and are described by the left (right) DQD detuning $\varepsilon_L$ ($\varepsilon_R$). Current will flow in a nanowire DQD through a series of downhill (in energy) single electron tunneling events (see level diagrams in Fig. 1B). In contrast with quantum well structures, current results from single electron tunneling events between electrically tunable zero-dimensional states in the DQD (*15, 22*). Electron tunneling results in microwave gain, which is accessed by measuring the transmission through the cavity (*16*).

To measure the gain, the cavity is driven with a coherent field at frequency $f_{in} = f_c$ with a power $P_{in}$. Measurements of the output power $P_{out}$ yield the power gain $G = C\, P_{out}/P_{in}$, where $C$ is a normalization constant set such that $G = 1$ when both DQDs are in Coulomb blockade (no current flow) (*16, 23*). With $V_{SD} = 0$, charge dynamics within the DQD result in an effective microwave admittance that damps the electromagnetic field inside of the cavity, yielding $G < 1$ (*18, 24, 25*). Application of a source-drain bias that drives sequential tunneling through the DQD can lead to gain in the cavity transmission, $G > 1$ (16). In Fig. 1D we plot $G$ as a function of $\varepsilon_L$ for $V_{SD} = 2$ mV and $f_{in} = f_c$. For downhill electron tunneling ($\varepsilon_L > 0$), we measure a maximum gain $G \approx 7$ (23). In contrast, for $\varepsilon_L < 0$, the left DQD can absorb a photon from the cavity, leading to loss $G \approx 0.2$ (18, 25). These data are acquired with the right DQD configured in Coulomb blockade such that the current is zero (15). For simplicity, we refer to a DQD as "on" when its detuning is set to achieve maximum gain and "off" when the DQD is configured in Coulomb blockade with $G = 1$.

We investigate the cavity response by measuring $G$ as a function of $f_{in}$ with $P_{in} = -120$ dBm, as shown in Fig. 2. The black curve is the "cold cavity transmission" obtained with both DQDs configured in the off state, where the maximum $G = 1$. Here the gain curve is a Lorentzian with a width set by the cavity decay rate $\kappa/2\pi = 2.6$ MHz. When $\varepsilon_L$ is set to the gain peak shown in Fig. 1D, we observe a maximum $G \approx 16$ at $f_{in} = 7880.30$ MHz. Similarly, with the right DQD on and the left DQD off, we observe a maximum $G \approx 6$ at $f_{in} = 7880.41$ MHz. In both configurations the observed gain is too small to reach the maser threshold. In contrast, the red curve in Fig. 2 shows the gain curve with both DQDs in the on state. Here the cavity response is sharply peaked at $f_{in} = 7880.25$ MHz, yielding a maximum gain $G \approx 1000$, which is much larger than the product of the individual gains.

We next examine the characteristics of the device in free running mode (with no cavity drive tone). Figure 3 shows the power spectral density $S(f)$ of microwave radiation emitted from the cavity in the on/on configuration. The spectrum is sharply peaked around $f = 7880.8$ MHz and has a full-width at half-maximum (FWHM) $\Delta_f = 34$ kHz, which corresponds to a coherence time $\tau_{coh} = 1/\pi\Delta_f = 9.4$ µs and a coherence length $l_{coh} = \tau_{coh}\ c = 2.8$ km, where $c$ is the speed of light. The measured linewidth is roughly a factor of 100 larger than the Schawlow-Townes prediction, but it is not uncommon for conventional semiconductor lasers to have broad emission linewidths (23, 26, 27). Time domain measurements of $\tau_{coh}$ are shown in (23).

The most striking evidence of above threshold maser action is obtained by comparing the statistics of the radiation emitted from the device in the off/on and on/on configurations (23). For this purpose we have sampled the voltages of the down-converted cavity output field to heterodyne detect the in-phase and quadrature phase components $I$ and $Q$ with a rate of 1 MHz after applying a 1 MHz digital filter. We store $4\times10^5$ individual $(I,Q)$ measurements in two-dimensional histograms $D(I,Q)$ to analyze their statistical properties. The measured $IQ$ histogram for the off/on configuration is shown in Fig. 4A. The histogram is centered near the origin and the extracted photon number distribution (Fig. 4B) is consistent with a thermal source (23). In contrast, Fig. 4C shows the $IQ$ histogram for the on/on configuration. Here the $IQ$ histogram has a donut shape, consistent with an above threshold maser (2). The extracted photon number distribution is peaked around a photon number $n = 8000$, giving strong evidence for above threshold behavior. The peak in the photon number distribution is well fit with a Gaussian lineshape, but its width is considerably larger than that of an ideal coherent state $\sqrt{\overline{N}} \approx 90$, where $\overline{N}$ is the average photon number (23). Time domain measurements of the maser emission indicate that charge noise fluctuations, which shift the detuning of the DQD gain medium, are most likely responsible for the broadening. Charge noise also occasionally shifts the system below threshold, leading to the small thermal component observed in Fig. 4D (23).

We have demonstrated a maser whose gain medium consists of electrically tunable semiconductor DQDs. Single electron tunneling in the DQDs provides the energy source for maser action and a maximum power gain of 1000 is observed. Above-threshold maser action is verified by measuring the statistics of the emitted photon field. Through further improvements in the cavity quality factor (*28*), it may be possible to exceed the lasing threshold with a single DQD emitter. In this case, theory predicts "thresholdless lasing" (*29*). Lastly, the large single particle level spacings allow for an operation frequency that is purely set by the cavity resonance frequency. This will enable maser operation across a very wide frequency range, spanning GHz to THz frequencies, a feature that is specific to gate defined quantum dots, where electron tunneling takes place between confined zero-dimensional electronic states.

**References and Notes:**
1. H. Haken, *Laser Theory*. (Springer, Berlin, 1983).
2. A. E. Siegman, *Lasers*. (University Science Books, Mill Valley, CA, 1986).
3. R. N. Hall, R. O. Carlson, T. J. Soltys, G. E. Fenner, J. D. Kingsley, *Phys. Rev. Lett.* **9**, 366 (1962).
4. M. I. Nathan, W. P. Dumke, G. Burns, F. H. Dill, G. Lasher, *Appl. Phys. Lett.* **1**, 62 (1962).
5. R. Dingle, W. Wiegmann, C. H. Henry, *Phys. Rev. Lett.* **33**, 827 (1974).
6. J. Faist *et al.*, *Science* **264**, 553 (1994).
7. Y. Yao, A. J. Hoffman, C. F. Gmachl, *Nature Photon.* **6**, 432 (2012).
8. D. Meschede, H. Walther, G. Müller, *Phys. Rev. Lett.* **54**, 551 (1985).
9. J. McKeever, A. Boca, A. D. Boozer, J. R. Buck, H. J. Kimble, *Nature (London)* **425**, 268 (2003).
10. T. Yoshie *et al.*, *Nature (London)* **432**, 200 (2004).
11. J. P. Reithmaier *et al.*, *Nature (London)* **432**, 197 (2004).
12. Z. G. Xie, S. Gotzinger, W. Fang, H. Cao, G. S. Solomon, *Phys. Rev. Lett.* **98**, 117401 (2007).
13. M. Nomura, N. Kumagai, S. Iwamoto, Y. Ota, Y. Arakawa, *Nat. Phys.* **6**, 279 (2010).
14. O. Astafiev *et al.*, *Nature (London)* **449**, 588 (2007).
15. W. G. van der Wiel *et al.*, *Rev. Mod. Phys.* **75**, 1 (2003).
16. Y.-Y. Liu, K. D. Petersson, J. Stehlik, J. M. Taylor, J. R. Petta, *Phys. Rev. Lett.* **113**, 036801 (2014).
17. A. Wallraff *et al.*, *Nature (London)* **431**, 162 (2004).
18. T. Frey *et al.*, *Phys. Rev. Lett.* **108**, 046807 (2012).
19. P.-Q. Jin, M. Marthaler, J. H. Cole, A. Shnirman, G. Schon, *Phys. Rev. B* **84**, 035322 (2011).
20. C. Fasth, A. Fuhrer, L. Samuelson, V. N. Golovach, D. Loss, *Phys. Rev. Lett.* **98**, 266801 (2007).
21. S. Nadj-Perge, S. M. Frolov, E. P. A. M. Bakkers, L. P. Kouwenhoven, *Nature (London)* **468**, 1084 (2010).
22. R. Hanson, L. P. Kouwenhoven, J. R. Petta, S. Tarucha, L. M. K. Vandersypen, *Rev. Mod. Phys.* **79**, 1217 (2007).
23. Supplementary materials are available on *Science* Online.
24. M. R. Delbecq *et al.*, *Phys. Rev. Lett.* **107**, 256804 (2011).
25. K. D. Petersson *et al.*, *Nature (London)* **490**, 380 (2012).
26. A. L. Schawlow, C. H. Townes, *Phys. Rev.* **112**, 1940 (1958).

27. P. W. Milonni, J. H. Eberly, *Laser Physics*. (John Wiley & Sons, New Jersey, 2010).
28. A. Megrant *et al.*, *Appl. Phys. Lett.* **100**, 113510 (2012).
29. I. Protsenko *et al.*, *Phys. Rev. A* **59**, 1667 (1999).

**Acknowledgements:**
We thank Karl Petersson for assistance with sample fabrication. Research supported by the Packard Foundation, the National Science Foundation (DMR-0819860 and DMR-0846341), DARPA QuEST (HR0011-09-1-0007), and the Army Research Office (W911NF-08-1-0189). Certain commercial equipment, instruments, or materials are identified in this paper to foster understanding. Such identification does not imply recommendation or endorsement by the National Institute of Standards and Technology, nor does it imply that the materials or equipment identified are necessarily the best available for the purpose. All data described in the paper are presented in this report and supplementary materials.

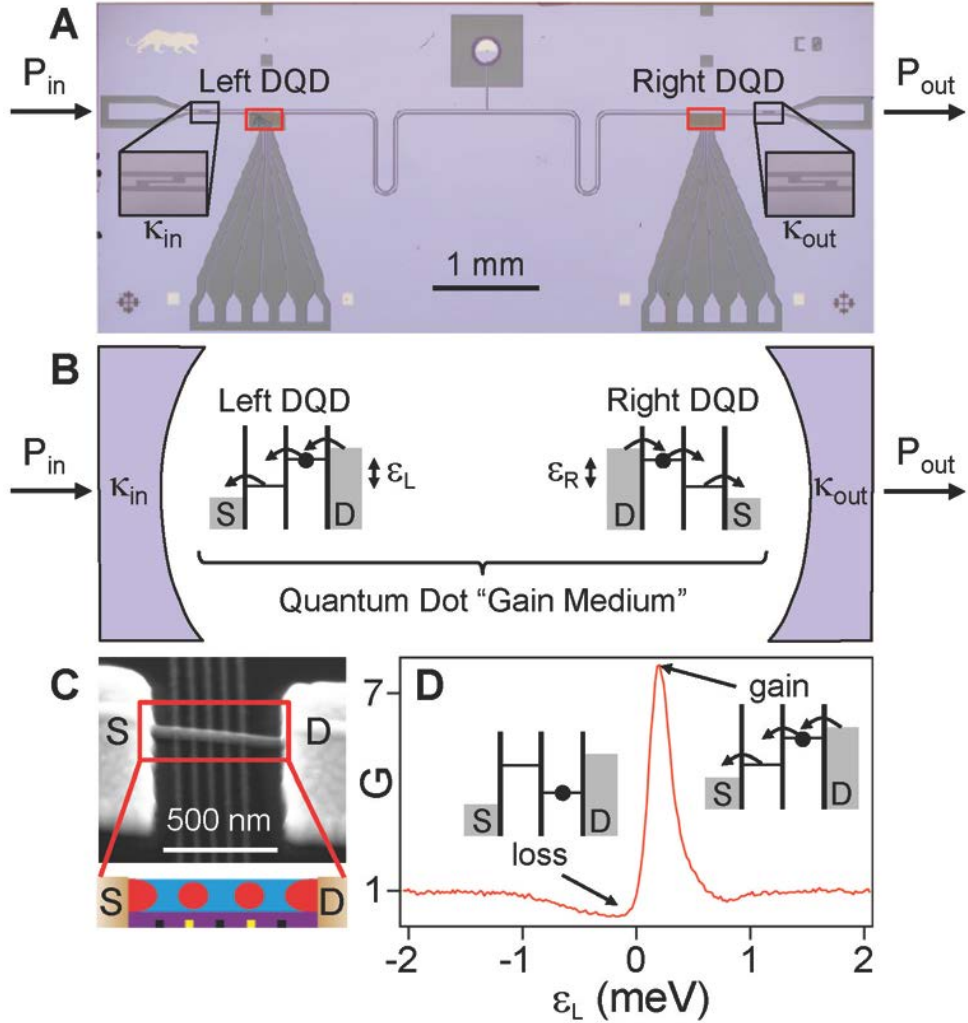

**Fig. 1. Double quantum dot micromaser.** (**A**) Optical micrograph of the DQD micromaser. Cavity photons are coupled to input and output ports with rates $\kappa_{in}$ and $\kappa_{out}$. (**B**) Schematic illustration of the DQD micromaser. Two DQDs are electric-dipole coupled to the microwave cavity. Single electron tunneling through the DQDs leads to photon emission into the cavity mode. Left (right) DQD detunings $\varepsilon_L$ ($\varepsilon_R$) are independently tunable. (**C**) Scanning electron microscope image of an InAs nanowire DQD. (**D**) $G$ as a function of $\varepsilon_L$ (measured at frequency $f_c$) with $V_{SD} =$ 2 mV and the right DQD configured in Coulomb blockade. Insets: For $\varepsilon_L > 0$ electron transport proceeds downhill in energy, resulting in a gain exceeding 7. With $\varepsilon_L < 0$ an electron will be trapped in the right dot until a photon is absorbed, resulting in cavity loss, $G < 1$.
6

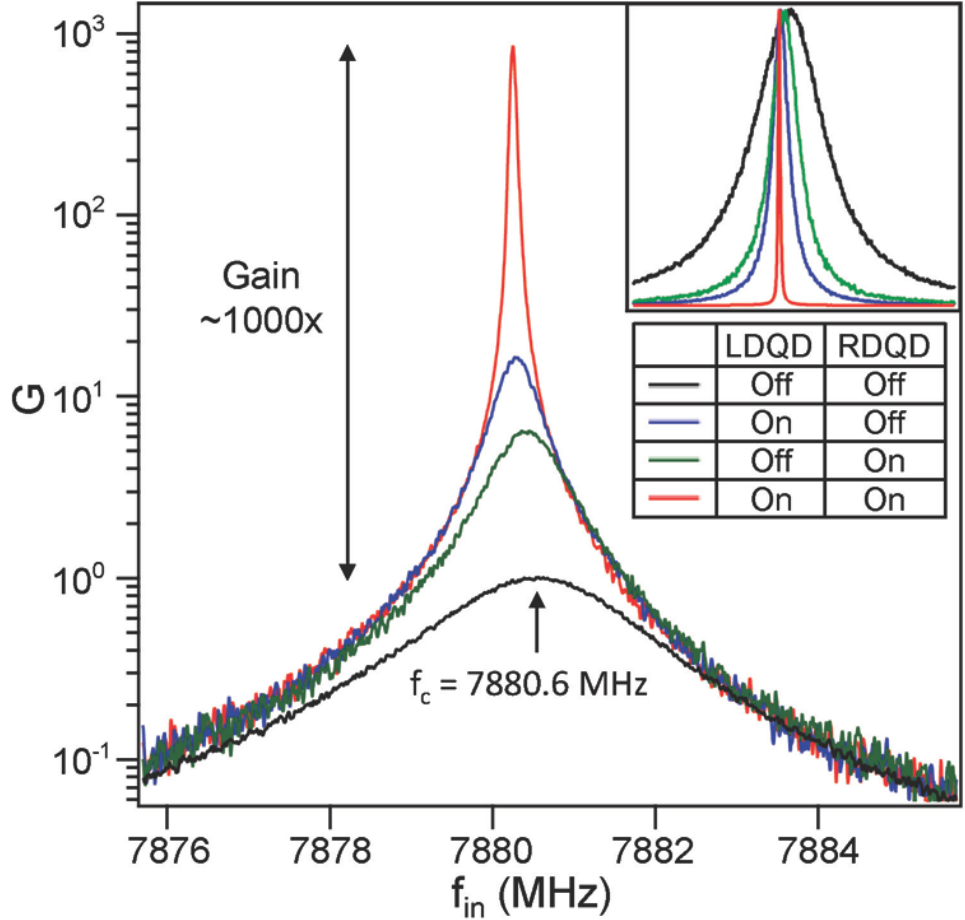

**Fig. 2. Microwave gain induced by single electron tunneling.** $G$ as a function of $f_{in}$ with $P_{in}$ = -120 dBm. The black curve is obtained with both DQDs in Coulomb blockade (in the off/off state). With the left DQD set at a detuning that results in gain (see Fig. 1D) and the right DQD in Coulomb blockade (on/off state), we measure a maximum $G \approx 16$. Similarly, in the off/on state we observe a gain of $\approx 6$. Maser action occurs when both DQDs are tuned to produce gain, resulting in $G \approx 1000$. Inset: Data plotted on a linear scale and normalized to the same height.

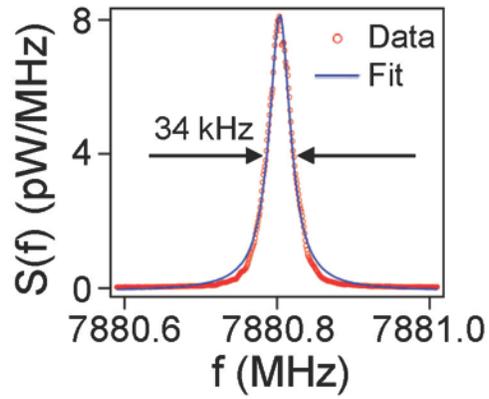

**Fig. 3. Maser coherence time.** Power spectral density $S(f)$ measured in free running maser mode (on/on state with no cavity drive applied). The maser emission peak width $\Delta_f = 34$ kHz (FWHM) yields a coherence length $l_{coh} = 2.8$ km.

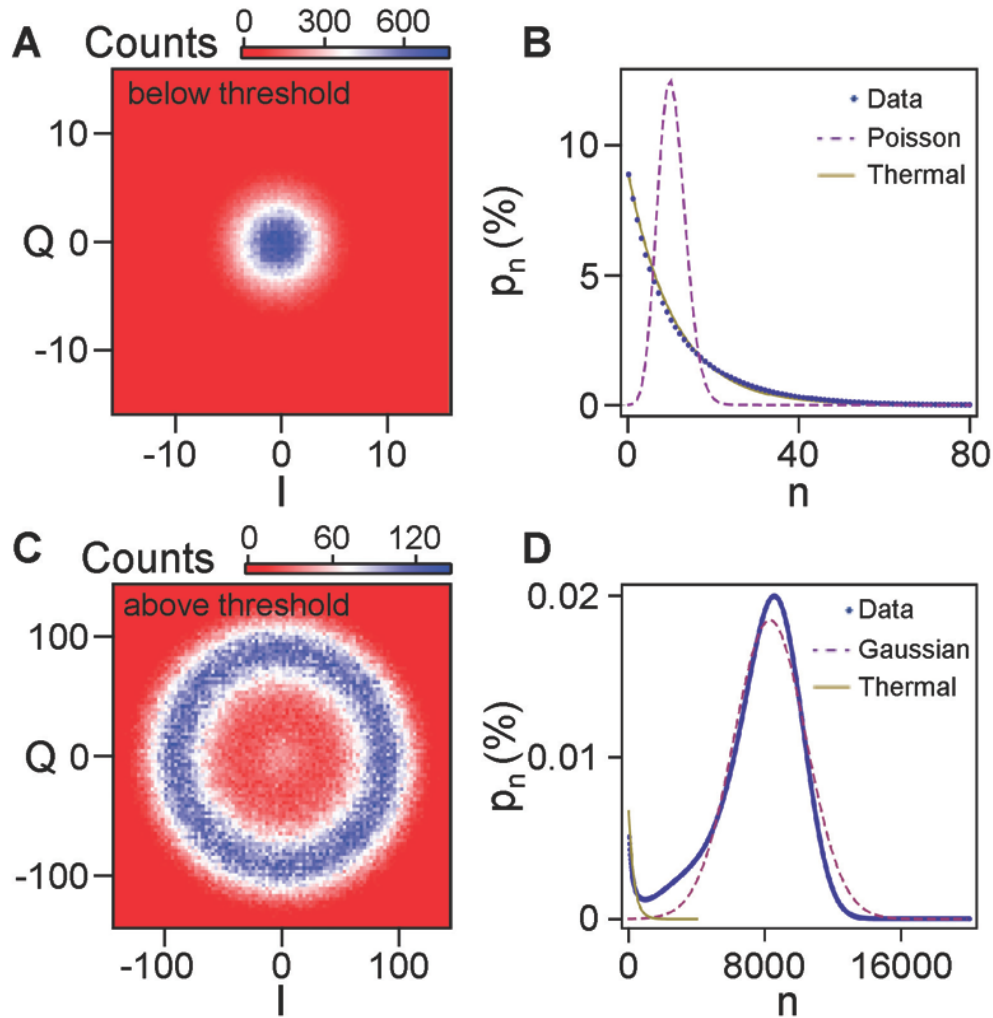

**Fig. 4. Photon statistics.** (**A**) *IQ* histogram acquired below threshold (off/on configuration). (**B**) The photon number distribution, $p_n$, extracted from the data in (A) is consistent with a thermal distribution (solid line). A Poisson distribution (dashed line) with $\overline{N} = 11.4$ is shown for comparison. (**C**) *IQ* histogram measured above threshold (on/on configuration). Here the extracted photon number distribution (**D**) is peaked around $n = 8000$ and is compared with a Gaussian distribution (dashed line). A small thermal component (solid line) is attributed to charge fluctuations, which shift the device below threshold.

# Supplementary Materials for

## Semiconductor double quantum dot micromaser

Y.-Y. Liu, J. Stehlik, C. Eichler, M. J. Gullans, J. M. Taylor, J. R. Petta

correspondence to: petta@princeton.edu

**This PDF file includes:**

Materials and Methods
Supplementary Text
Figs. S1 to S6
References (30–39)

**Materials and Methods**

1. Device Structure

The device consists of a cavity and two double quantum dots (DQDs) as shown in Fig. S1A. The cavity is a half-wavelength ($\lambda/2$) superconducting Nb transmission line resonator with a bare resonance frequency $f_c$ = 7880.55 MHz. The electric field profile of the fundamental mode is shown in Fig. S1A. There is an electric field node at the center of the cavity and anti-nodes near the input and output coupling ports of the cavity. For small cavity photon numbers $n \approx 1$, the electric field at the antinode of the cavity is estimated to be 0.4 V/m (*30*). The input and output ports of the resonator are coupled to 50 ohm transmission lines with characteristic rates $\kappa_{in}/2\pi = \kappa_{out}/2\pi = 0.39$ MHz. Internal losses are described by the internal decay rate $\kappa_{int}$. Therefore the total decay rate $\kappa = \kappa_{in} + \kappa_{out} + \kappa_{int}$. The cavity linewidth (FWHM) is $\kappa/2\pi = 2.6$ MHz, giving a loaded cavity quality factor $Q_c \approx 3000$.

The gain medium consists of two InAs nanowire DQDs (*25*). Each DQD is fabricated by placing an InAs nanowire over five Ti/Au depletion gates (*20, 21*). The gate voltages are generated using low-pass filtered digital-to-analog converters. As schematically shown in Fig. S1B, the left DQD is formed by biasing gates $B_{L1}$, $L_1$, $C_1$, $R_1$, $B_{R1}$ and the right DQD is formed by biasing gates $B_{L2}$, $L_2$, $C_2$, $R_2$, $B_{R2}$. Gate voltages are used to tune the chemical potential of each quantum dot (*15*). Here $\mu_{L1}$, $\mu_{R1}$ are the chemical potentials of the left and right dots for the left DQD. Similarly $\mu_{L2}$, $\mu_{R2}$ are the chemical potentials of the left and right dots for the right DQD. The left DQD detuning $\varepsilon_L = \mu_{R1} - \mu_{L1}$ and right DQD detuning $\varepsilon_R = \mu_{L2} - \mu_{R2}$ are independently tunable. A source-drain bias voltage $V_{SD}$ is applied to both DQDs through an on-chip three-turn spiral inductor that is connected to the voltage node of the cavity (*25*). This sets $\mu_S$ and $\mu_D$, the Fermi level of the source and drain. We measure the resulting total current through the two DQDs using a DL instruments current preamplifier as shown in Fig. S2.

The semiconductor DQDs are coupled to the cavity through the electric dipole interaction. The cavity electric field $E$ couples to the electric dipole moment $d \approx 1000\ ea_0$ of a single excess electron trapped in the DQD (*25*). Here $e$ is the electronic charge and $a_0$ is the Bohr radius. To enhance the electric field at the position of the DQD and maximize the charge–cavity coupling rate, the source contact to each nanowire is lithographically connected to a voltage antinode of the cavity and the drain contact of each nanowire is connected to the resonator ground plane (see Fig. S1). Through standard microwave characterization, we measure charge-cavity coupling rates $g_1/2\pi \sim g_2/2\pi \sim 30$ MHz for each DQD (*16, 18, 25*). These values are consistent with rates obtained in other quantum dot systems (*18, 24, 31, 32*).

2. Microwave setup

A complete microwave circuit diagram is shown in Fig. S2. Four different types of measurements are performed in the experiment: 1) Measurements of the cavity power gain $G$ as a function of left DQD detuning $\varepsilon_L$ using homodyne detection (Fig. 1D of main text), 2) Measurements of the cavity power gain $G$ as a function of frequency using a network analyzer (Fig. 2 of main text), 3) Measurements of the power spectral density of radiation emitted from the maser in the absence of a drive tone (Fig. 3 of main text), and 4) Heterodyne measurements of the free-running maser emission (Fig. 4 of main text). These different measurement configurations are shown in different colors in Fig. S2 (blue for measurement 1, green for measurement 2, red for measurement 3, and gold for measurement 4). Wiring that is common to all of these measurements is shown in black. In the following we describe each measurement configuration in detail.

## 2.1 Measurements of the cavity power gain as a function of detuning

Figure 1D of the main text shows the power gain $G$ as a function of left DQD detuning $\varepsilon_L$, with the right DQD configured in Coulomb blockade. The power gain is measured by driving the cavity at frequency $f_{in}$ and power $P_{in}$. Here $f_{in}$ is chosen to match the bare cavity resonance frequency $f_c$. The cavity output signal (with power $P_{out}$) is amplified with a high electron mobility transistor (HEMT) amplifier, which is mounted at the 4K stage of the dilution refrigerator. Further amplification is performed at room temperature, followed by homodyne demodulation (blue path in Fig. S2). Here the signal exiting the cryostat is mixed with frequency $f_{in}$ using a Marki Microwave IQ0618LXP IQ mixer. The signal from the $I$ and $Q$ quadratures are filtered by SBLP-39+ low pass filters and then amplified by SRS SR560 preamplifiers configured with a 10 kHz low frequency cutoff. The resulting analog signals are digitized with a NI DAQ PCI-6221. The power gain is defined as $G = C\, P_{out}/P_{in}$, where $C$ is a normalization constant set such that $G = 1$ when both DQDs are in Coulomb blockade (no current flow) (*16, 25*).

## 2.2 Measurements of the cavity power gain as a function of frequency

The data presented in Fig. 2 of the main text are also acquired using homodyne detection. Here demodulation is achieved using an Agilent E8363B network analyzer (green path in Fig. S2).

## 2.3 Measurements of the power spectral density

Emission from the cavity is always measured in the absence of a cavity drive tone. The data shown in Fig. 3 of the main text are acquired by turning off the cavity drive ($f_{in}$) and measuring the cavity emission using an Agilent E4405B spectrum analyzer. This measurement configuration is indicated by the red path in Fig. S2.

## 2.4 Heterodyne measurements of the free-running maser emission

The data presented in Fig. 4 are obtained using heterodyne detection. The cavity drive tone $f_{in}$ is turned off for these measurements. Radiation generated by the cavity is first passed through a chain of circulators, to prevent noise from the amplification chain from populating the cavity. The signal then reflects off of a Josephson parametric amplifier (JPA), which is thermalized at the base temperature of the dilution refrigerator. In the absence of a JPA pump field the signal will be reflected off of the JPA with unity gain. In the presence of a JPA pump field, the signal will be reflected with power gains ranging from 0 to 30 dB, depending on the pump power. The reflected signal is then further amplified by the 4K HEMT amplifier and room temperature amplifiers. After amplification the output is first down-converted to an intermediate frequency (IF) of 12.5 MHz, by mixing it with a local oscillator tone at frequency $f_{lo} = f_c + 12.5$ MHz. A separate 12.5 MHz reference tone is generated by mixing $f_{lo}$ and $f_{ref}$, where $f_{ref} = f_c$. The down-converted signal is then filtered by two 23 MHz low-pass filters and amplified by SR445A preamplifier before being digitized by an Alazar ATS9625 FPGA/digitizer board at 250 MHz. The FPGA further demodulates the signal into $I$ and $Q$ quadratures by multiplying the digitized signal with the reference tone (resulting in the $I$ signal) and with a 90 degree phase shifted reference tone (resulting in the $Q$ signal). The resulting $I$ and $Q$ waveforms are then filtered by a 12.5 MHz boxcar FIR filter to remove the remaining carrier and 25 MHz signals and then passed to the computer. A further 1 MHz boxcar filter is applied inside the computer. All the coherent sources and the acquisition card are phase locked to each other using a 10 MHz reference signal generated by a Rubidium clock.

For below threshold measurements (Fig. 4A and B of the main text) the signal is weak and we therefore amplify it using the JPA. We detune the frequency of the pump field $f_{pump}$ from the cavity frequency by $f_{pump} - f_c = -6$ MHz to achieve phase-preserving amplification. The JPA pump field is coupled into the system using a directional coupler (Krytar 120420). We interferometrically cancel out the reflected pump tone, which is unwanted in our detection chain, by adding a 180 degree phase shifted tone at the same frequency $f_{pump}$ through a second directional coupler port. Using a small magnetic field produced by a coil of superconducting cable near the JPA and by varying power of the pump field, we configure the JPA for +19 dB gain. For above threshold measurements presented in Fig. 4C and D of the main text the cavity emission is high and consequently measurements are performed without the JPA. This is achieved by leaving the pump field turned off.

### 3. Noise Background Calibration

The detected power spectral density, expressed in units of photons per Hz per second, is given by $N(f) = G_{det} [N_{offset} + N_{vac} + G_{jpa} \eta_{col} S(f)/hf]$. Here, $G_{det}$ is the effective gain of the entire detection chain after the JPA, $N_{offset}$ is the effective noise number as seen from the output of the JPA, $G_{jpa}$ is the gain of the JPA and $N_{vac} \approx G_{jpa}$ is the noise originating from vacuum fluctuations amplified by the JPA. As defined in the main text $S(f)$ is the total power spectral density emitted from the cavity. The collection efficiency $\eta_{col} = \eta_{cable} \eta_{cavity} = 0.065$ is limited by radiation loss 1-$\eta_{cable} \approx 0.43$ between the cavity and the JPA, and by the finite cavity emission efficiency $\eta_{cavity} = \kappa_{out}/\kappa \approx 0.15$. Thus only 6.5% of the total power emitted from the cavity reaches the JPA.

To estimate the noise offset $N_{offset}$ we measure the power spectral density $N(f)$ when the JPA is turned on and when the JPA is turned off. For both these measurements we have no cavity drive applied and the DQDs are biased in the off/off configuration. The cavity emission spectrum $S(f)$ in both cases is thus approximately zero up to a small residual thermal background. Comparing the two power spectral density $N(f)$ measurements with the known JPA gain of 19 dB and using the above formula for $N(f)$, we obtain $N_{offset} \approx 225$, see Eichler *et al.* (*33*). We estimate the error of this calibration to be about ±10%. For later use we also determine the effective noise number $N_{noise} \approx (2N_{offset} + G_{jpa})/G_{jpa}$, corresponding to the noise level when referencing back to the input of the JPA and when measuring with the heterodyne detection chain.

**Supplementary Text**

The data in Fig. 3 and 4 of the main text were taken after a JPA was installed in the cryostat. As a result, the sample had to be thermally cycled in the dilution refrigerator.

### 1. Maser Coherence

We determine the coherence length and time of the maser using two complementary sets of measurements. In the first measurement, we simply measure the width of the free running maser emission peak (Fig. 3 main text). We compare this result with time domain measurements of the emitted microwave signal in free running maser mode (Fig. S3B).

### 1.1 Power spectral density measurements

We first extract the coherence length from the measured power spectral density $S(f)$, which is shown in Fig. S3A. Measurements of $S(f)$ are described in the Materials and Methods. The emission peak is nicely fit by a Lorentzian lineshape $S(f) = \frac{S_0 \Delta_f^2}{4(f-f_0)^2 + \Delta_f^2}$. From this fit we extract the emission center frequency $f_0 = 7880.8$ MHz, the peak density $S_0 = 8$ pW/MHz, and the FWHM $\Delta_f = 34.2 \pm 0.2$ kHz. Using standard definitions, this corresponds to a coherence time $\tau_{coh} = 1/\pi\Delta_f = 9.4$ μs and a coherence length $l_{coh} = \tau_{coh}$ c $= 2.8$ km (*27*).

### 1.2 Time-domain measurements of the coherence time

The linewidth measurements are supported by time-domain analysis of the signal emitted from the cavity. We measure the first-order autocorrelation function $g^{(1)}(\tau) = |[ \Sigma\ V(t)\ V^*(t+\tau) ]|$, where $V(t) = I(t) + iQ(t)$ and $V^*$ its complex conjugate. We configure the DQDs in masing mode and record $V(t)$ with a bandwidth of 5 MHz (much greater than the 34 kHz emission linewidth). The autocorrelation function is normalized such that $g^{(1)}(\tau=0) = 1$ and fit with the expression $\exp[-(\tau/\tau_{coh})^p]$ with $p = 1.45$ and $\tau_{coh} = 14.6$ μs (see Fig. S3B).

### 1.3 Comparison to Schawlow-Townes formula

Above the masing threshold the Schawlow-Townes formula predicts a relation between the total output power of the maser $P_{tot}$ and its minimal linewidth $\Delta_f$, such that $\Delta_f = \pi h f_c (\kappa/2\pi)^2/P_{tot}$ (*26*). We obtain the total cavity output power by integrating the power spectral density shown in Fig. S3A as $P_{tot} = \int df\ S(f) = 3.6 \times 10^{-13}$ W. This corresponds to a Schawlow-Townes limit for the masing linewidth of $\Delta_f = 0.4$ kHz, which is two orders of magnitude smaller than the measured linewidth. We attribute this deviation to charge fluctuations in the DQDs.

## 2. Photon Statistics

To unambiguously characterize the masing behavior of the DQD device we have analyzed the photon statistics of the radiation emitted from the cavity. For this purpose we have recorded two-dimensional histograms $D(I,Q)$ from $4\times10^5$ individual $(I,Q)$ measurements, which have been sampled at a rate of 1 MHz after applying a 1 MHz digital filter. The measurement system is analyzed by recording histograms in the off/off configuration to first determine the contribution from amplifier noise. Then we examine histograms below threshold (off/on) and above threshold (on/on). The main result is that the off/on case is described by a thermal distribution, while the on/on configuration has a dramatically different *IQ* histogram that more closely resembles the Poisson distribution expected for a maser.

### 2.1 Amplifier background

In a first experiment we left the cavity in the vacuum state by configuring the device in the off/off state and recording a reference histogram $D(I,Q)$ of the background noise of the amplification chain only, see Fig. S4A. The JPA gain in this and the following measurement has been set to 19 dB. The *I* and *Q* axes of the histograms are scaled such that the average photon number calculated from the off/off histogram $\int D(I,Q)(I^2 + Q^2)dIdQ$ equals the effective noise number $N_{noise}$ of the detection chain when referenced back to the input of the JPA (*34*). The noise number has been extracted from the power spectral density measurements as described above and is equal to $N_{noise} = 6.6$ for a JPA gain of 19 dB. The histogram data is very well described by a Gaussian distribution $\sim \exp\left[-\frac{(I^2+Q^2)}{2\sigma^2}\right]$ with a width of $\sigma = \sqrt{N_{noise}/2}$ as expected for thermal amplifier noise, see Fig. S4B for a fit of the histogram data to a Gaussian function.

## 2.2 Below threshold (off/on state)

In a second experiment we have turned on only the right DQD, leaving the left DQD in Coulomb blockade. In this configuration, the radiation emitted by the right DQD results in a broader histogram as shown in Fig. S4C. The histogram data, again, are fit very well using a Gaussian distribution with a width of $\sigma = \sqrt{(N_{noise} + \bar{N})/2}$, where the additional $\bar{N} \approx 11.4$ photons arise from the DQD induced cavity emission, see Fig. S4D. With the chosen scaling of the $I$ and $Q$ axes, $\bar{N}$ corresponds to the average number of photons arriving at the JPA input in a mode determined by the filter band.

Based on the iterative maximum likelihood approach we have used the measured *IQ* histogram to reconstruct the mostly likely photon number distribution $p_n$ of the radiation arriving at the JPA before the amplifier noise is added (*35*). The reconstructed distribution is shown in Fig. S4E and is very close to a thermal distribution $p_n = (\bar{N}/(\bar{N} + 1))^n/(\bar{N} + 1)$, (solid line). In contrast, the data are poorly fit to a Poisson distribution $p_n = \frac{\exp(-\bar{N})\bar{N}^n}{n!}$ (dashed line). This indicates that the system in this case is below the maser threshold. The average thermal photon number of $\bar{N} \approx 11.4$ corresponds to an effective temperature of $T = 4.5$ K for the radiation in the detected mode, which is comparable to the noise temperature of a HEMT amplifier.

## 2.3 Above threshold (on/on state)

For the next histogram measurement we have turned on both DQDs. Due to the much larger average emission rate, corresponding to $\bar{N} \approx 7500$ in the detected mode, we have turned off the JPA for this measurement to avoid non-linear effects resulting from amplifier saturation. Turning off the JPA results in a higher noise number of $N_{noise} \approx 450$ and correspondingly to a different scaling of the histogram axes. The resulting histogram shown in Fig. S4F appears like a circle that has been convolved with a Gaussian. As described in Siegman this particular phase space distribution is expected for coherent radiation, where the radius $A_e \approx 88$ of the broadened ring corresponds to the average coherent amplitude of the radiation field (*2*). The width of the ring $\sigma_e = \sqrt{\frac{N_{noise}+2\delta A^2}{2}} \approx 19$, is determined by the detection noise $N_{noise}$ and by residual fluctuations in the coherent amplitude $\delta A^2 \approx 135$. In addition to the ring-like distribution we observe a finite population in the center of the histogram. We attribute this to charge fluctuations, which randomly bring the system into the sub-threshold state. We therefore fit a cut through the histogram data to a weighted sum of a Gaussian function with width of $\sigma \approx 34$ and a "donut shaped" function $C(I,Q)$ = $\int_0^{2\pi} \frac{d\theta}{2\pi}\exp(-[I + A_e\cos(\theta)]^2/2\sigma_e^2 - [Q + A_e\sin(\theta)]^2/2\sigma_e^2)$, see Fig. S4G. We confirm the coherent photon statistics of the masing emission by reconstructing the photon number distribution, see Fig. S4H. For perfectly coherent maser emission we would expect to observe a shot noise limited Poisson distribution, which in the limit of large amplitude $A_e \gg 1$ is well approximated by a Gaussian distribution of the form $p_n \sim exp\left[-\frac{(n-A_e^2)^2}{2A_e^2}\right]$. However, due to the charge noise induced amplitude fluctuations $\delta A^2$ we find the reconstructed distribution more closely resembles a broadened Gaussian distribution $p_n \sim exp\left[-\frac{(n-A_e^2)^2}{2A_e^2(1+4\delta A^2)}\right]$ (dashed line). Here, we have used the same values for $\delta A^2$ and $A_e$ as extracted from the fit to the histogram, which demonstrates consistency between the two types of analysis. Apart from the broadening effect, the observed peak in the number distribution at around $n \approx A_e^2$ is a clear indication for coherent maser emission. The enhanced distribution close to $n = 0$ originates from the thermal radiation (solid line) emitted when the system is below the masing threshold. From the relative weights between the thermal distribution (solid line) and the Gaussian distribution (dashed line) we estimate that the system spends about 2% of the time below threshold.

2.4 Role of charge noise fluctuations

Charge noise fluctuations are known to lead to dephasing in charge and spin qubits fabricated in semiconductor systems (*16, 36*). Figure S5 shows a time series of the maser emission in free-running mode, *I(t)*. The maser oscillations are interrupted by abrupt charge switching events that switch the device into the sub-threshold state. We anticipate that further improvements in the maser linewidth can be achieved by reducing the amount of charge noise present in the system. One path forward is to replace the InAs nanowire quantum dots with quantum dots formed in buried two-dimensional electron gases, which are known to have approximately four times less charge noise (*37*).

2.5 Attempted measurement of maser behavior through threshold

Figure 4 of the main text clearly shows a qualitative difference between the photon statistics obtained above and below the maser threshold. We attempted to probe the dynamics of the device near the masing threshold by measuring the photon emission rate $\Gamma_{out}$ as a function of power gain *G* (see Fig. S6A). The emission rate displays a dramatic increase near $G \sim 100$. There is a significant amount of scatter in the data due to charge noise fluctuations in the device, which make a quantitative analysis of threshold behavior impractical at this point in time. In addition, the emission peak width $\Delta_{3\,dB}$ is plotted as a function of *G* in Fig. S6B. The data nicely follow the expected gain-bandwidth relation $\Delta_{3\,dB} \propto \kappa/\sqrt{G}$ over two decades of *G*. Scatter in the data are due to charge noise fluctuations. The origin of systematic deviations from the expected scaling for small and large gains remains to be investigated in future studies.

3. Theory of phonon-assisted gain

Since the maser is operating in the high temperature regime, a reasonable description of the dynamics can be found from standard maser equations for the average population inversion of each dot *N* and the field amplitude $\alpha$ (whose magnitude is proportional to the square root of the photon number). In the presence of an external drive $\Omega$ (*1, 19, 38, 39*):

$$\dot{\alpha} = -\left(\frac{\left[\kappa - \frac{\chi(\varepsilon)N}{N_0(\varepsilon)}\right]}{2} + i\delta\right)\alpha + \Omega, \qquad (1)$$

$$\dot{N} = \Gamma_p(\varepsilon)[N_0(\varepsilon) - N] - \chi(\varepsilon)N|\alpha|^2. \qquad (2)$$

Here $\kappa$ is the cavity linewdith, $\chi$ is the combined gain rate of each emitter, $\delta = \omega_d - \omega_c$ is the detuning of the driving field at frequency $\omega_d$ from the cavity frequency $\omega_c$ (where we absorbed the cavity pull from the maser into this definition), $\Gamma_p$ is the repumping rate arising from the finite source drain bias, and $N_0$ is the steady state population inversion in the absence of the cavity. We have made explicit the dependence of these parameters on the detuning $\varepsilon$ between the left and right dots in each DQD.

When $\chi < \kappa$, the system is below threshold and we can neglect the saturation term in Eq. 2 so that $N = N_0$. In this case, the gain is given by the simple formula

$$G(\delta, \varepsilon) = \frac{\kappa^2}{[\kappa - \chi(\epsilon)]^2 + 4\delta^2}. \qquad (3)$$

By measuring the peak gain in each dot as a function of the detuning $\varepsilon$, as in Fig. 1D of the main text, we can extract the gain rate $\chi(\varepsilon)$. For both dots we find the peak gain rate is $\sim 2\pi \cdot 2$ MHz. This implies each dot is individually below threshold, but when both dots are tuned to maximum gain, the total gain rate is greater than the cavity decay $\kappa = 2\pi \cdot 2.6$ MHz and we expect the system to begin masing, consistent with our observations.

Above threshold, the photon number in the cavity saturates. This is also predicted from Eq. 1 and 2 where, in the absence of a drive, the steady state solution for the photon number is

$$|\alpha|^2 = \frac{\Gamma_p(\epsilon)}{\chi(\varepsilon)} \frac{\chi(\varepsilon) - \kappa}{\kappa}. \qquad (4)$$

Since charge noise leads to noise in the detuning $\varepsilon$, it is clear from this equation that charge noise will lead to noise in the output photon number distribution as discussed in the main text. However, since the maser is operating with only two emitters, to accurately describe the dynamics above threshold we need to take into account quantum fluctuations as well. A detailed theoretical model of these effects, including the microscopic origin of the gain, will be the subject of future work.

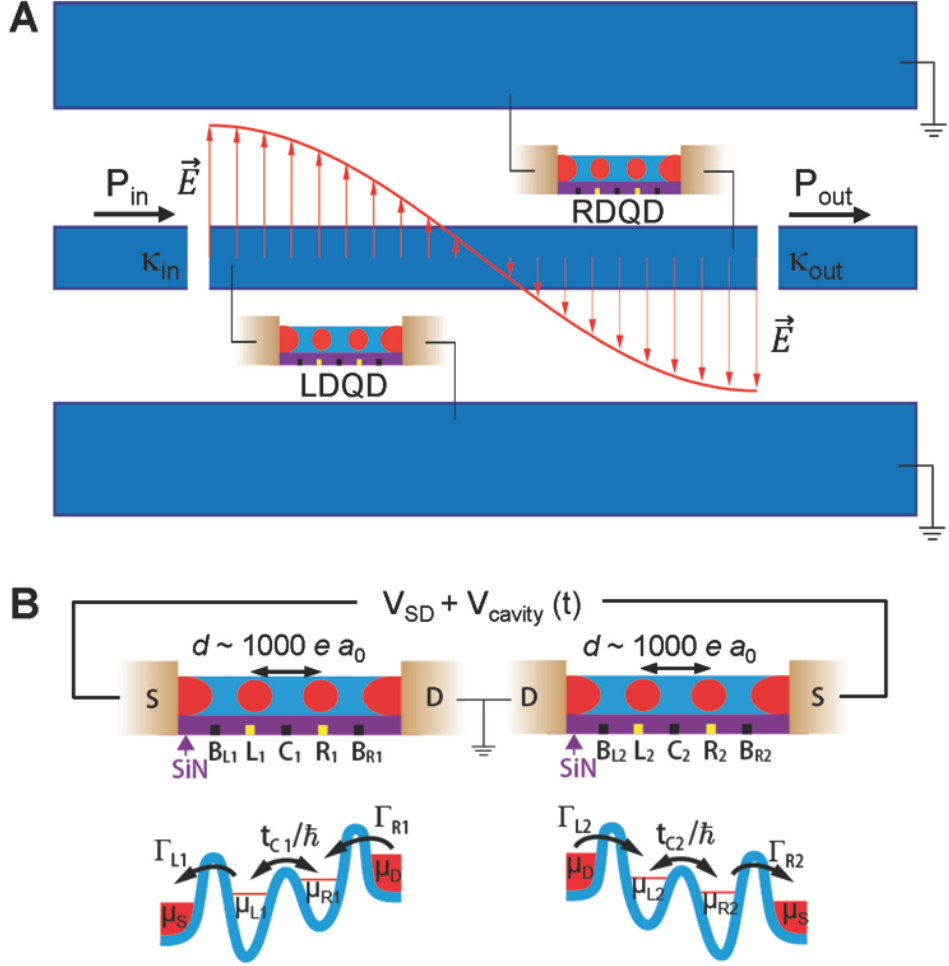

**Fig. S1.**
(**A**) Schematic illustration of the device. The electric field profile of the superconducting resonator is shown. The resonator is coupled to a 50 ohm transmission line by input (output) coupling capacitors that are described by rates $\kappa_{in}$ ($\kappa_{out}$). Cavity power gain $G = C\, P_{out}/P_{in}$, where $C$ is a normalization constant set such that $G = 1$ when both DQDs are in Coulomb blockade (no current flow). To maximize coupling between the DQD and the cavity field, the drain contact of each nanowire is physically contacted to the ground plane of the resonator, while the source contact of each nanowire is physically contacted to an anti-node of the resonator. (**B**) The left (right) DQD is formed by biasing gates $B_{L1}$, $C_1$, $B_{R1}$ ($B_{L2}$, $C_2$, $B_{R2}$) at negative voltages to form left and right tunnel barriers with rates $\Gamma_{L1}$ and $\Gamma_{R1}$ ($\Gamma_{L2}$ and $\Gamma_{R2}$), and an interdot tunnel barrier with tunnel rate $t_{C1}/\hbar$ ($t_{C2}/\hbar$). A source-drain bias $V_{SD} = V_S - V_D = (\mu_D - \mu_S)/|e|$ is applied to both DQDs. The DQDs are electric-dipole coupled to the microwave resonator. The dipole moment $d \approx 1000\, e\, a_0$, where $e$ is the electronic charge and $a_0$ is the Bohr radius.

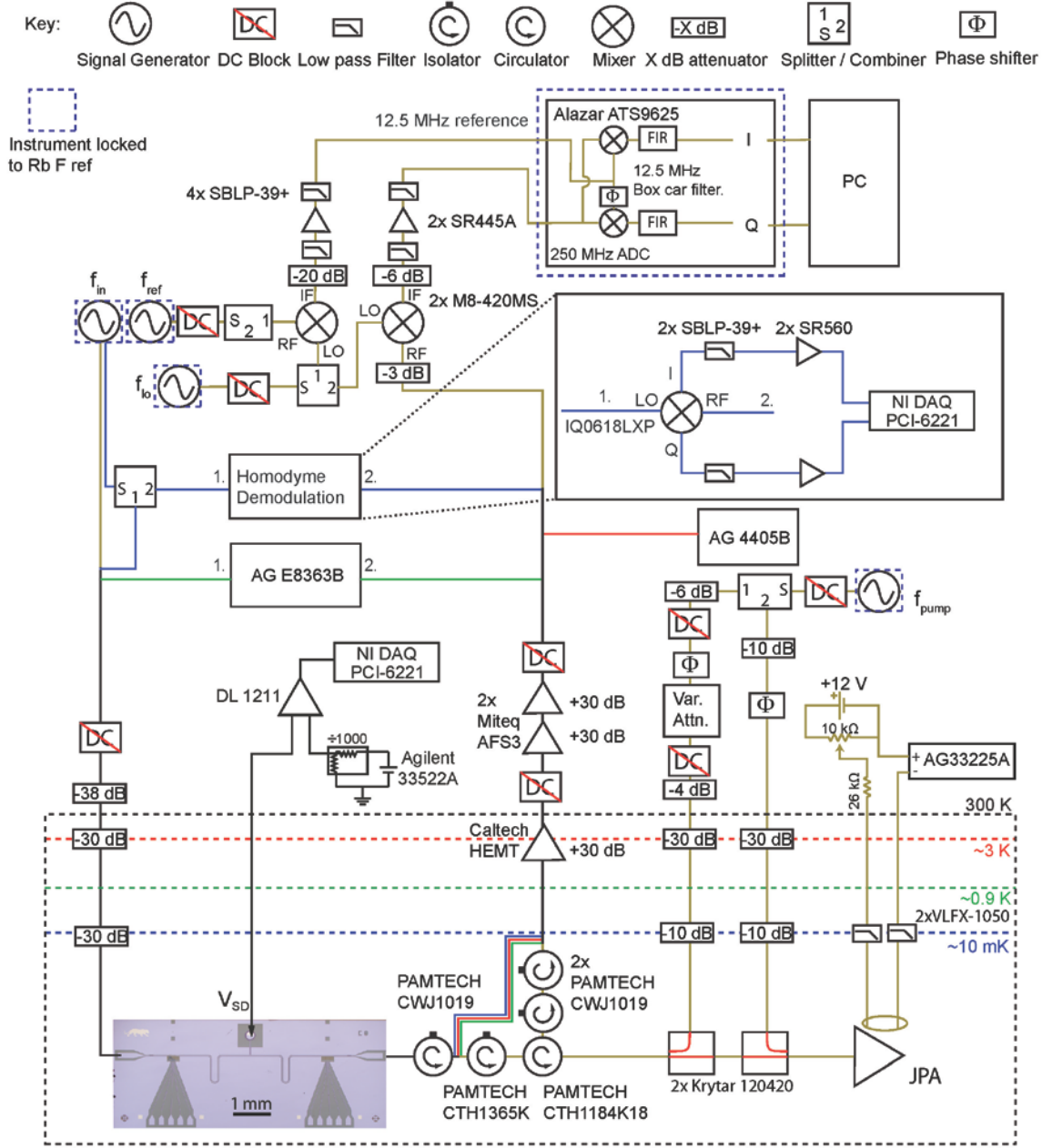

**Fig. S2**

Schematic diagram of the measurement setup. Four different measurement setups are used. Connections specific to each setup are color coded (blue, green, red, and gold), while connections common to all setups are shown in black.

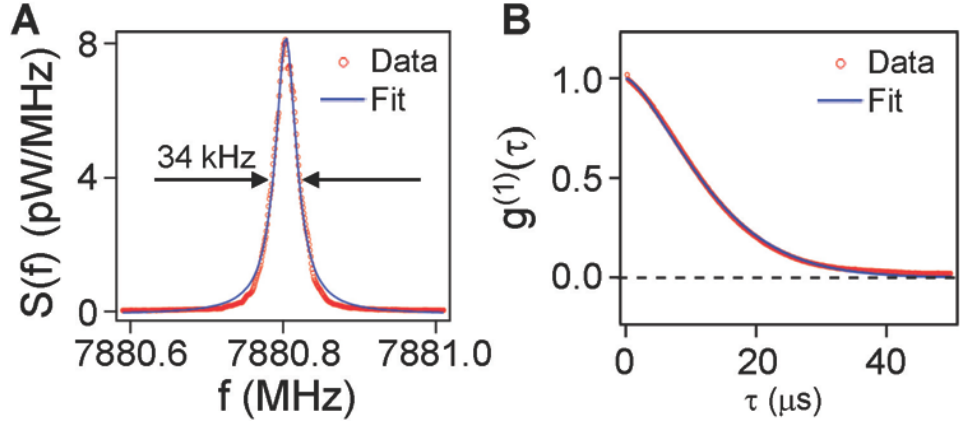

**Fig. S3**
(**A**) Power spectral density $S(f)$ measured with a resolution bandwidth of 1 kHz (dots) and fit to a Lorentzian (solid line). The full-width at half-maximum $\Delta_f = 34$ kHz, corresponds to a coherence time $\tau_{coh} = 1/\pi\Delta_f = 9.4$ μs and a coherence length $l_{coh} = \tau_{coh}\, c = 2.8$ km. (**B**) Measured autocorrelation function $g^{(1)}(\tau)$ (dots) extracted from a time trace of the $I$ and $Q$ quadratures and fit (blue). The blue solid line is a fit to the form $g^{(1)}(\tau) = e^{-(\frac{\tau}{\tau_{coh}})^p}$ with $p = 1.45$ and $\tau_{coh} = 14.6$ μs.

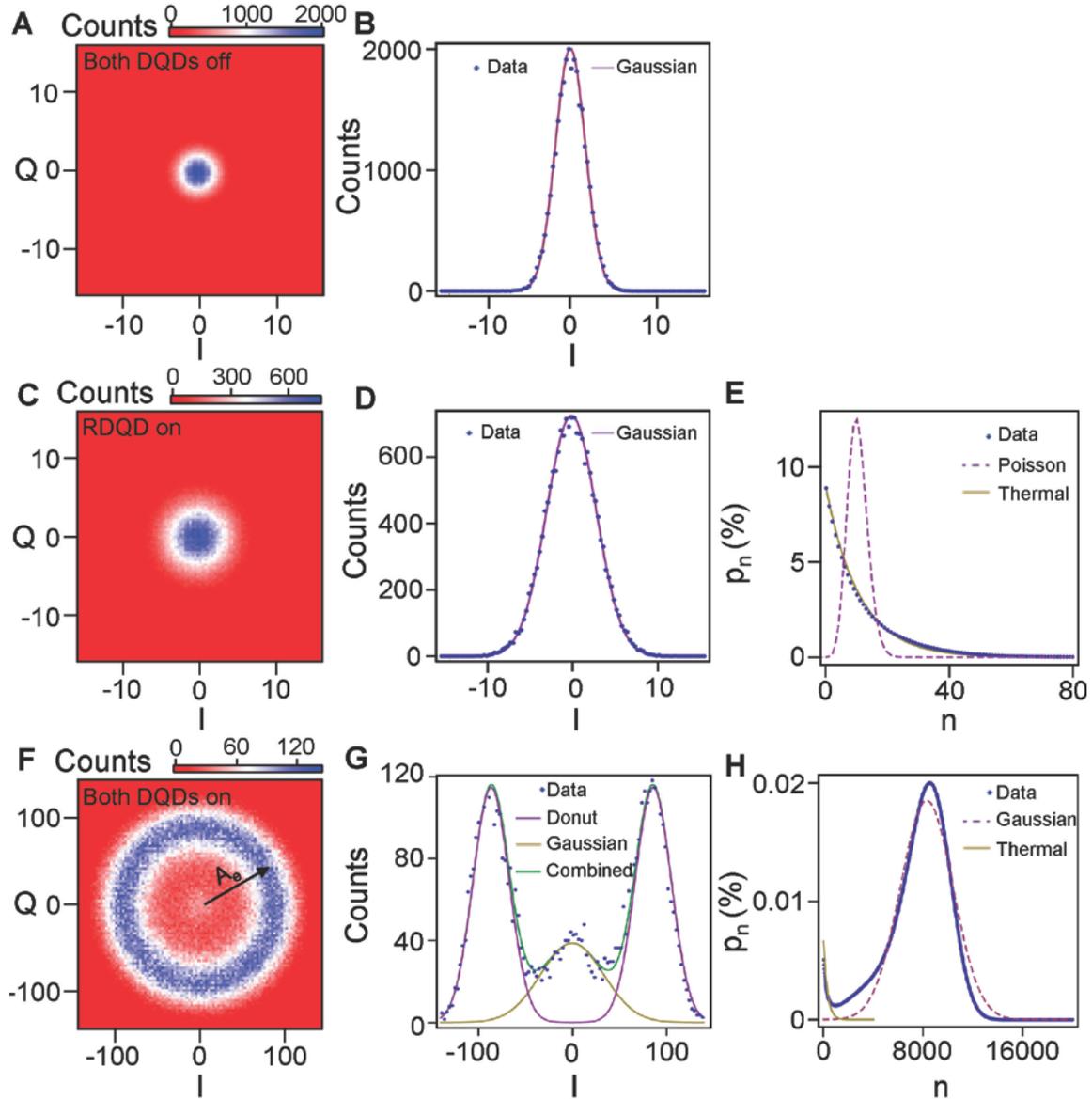

**Fig. S4**
(**A**) Histogram data taken with a JPA gain of 19 dB when both DQDs are Coulomb blockade. (**B**) Cut through the histogram in (A) at $Q = 0$ and a Gaussian fit. (**C**) Histogram taken with a JPA gain of 19 dB with the right DQD turned on. (**D**) Cut through histogram in (C) at $Q = 0$ and a Gaussian fit. (**E**) Photon number distribution of the RDQD emission extracted from the histogram in (A) and (C). Fit to a thermal distribution (solid line) and a Poisson distribution (dashed line) for comparison. (**F**) Histogram measured with both DQDs turned on. (**G**) Cut through the histogram in (F) with individual fits to a Gaussian (yellow), to a donut shaped function (purple) and a combined fit (green). (**H**) Photon number distribution extracted from (F) compared to a Gaussian distribution (dashed line) and a thermal distribution (solid line).

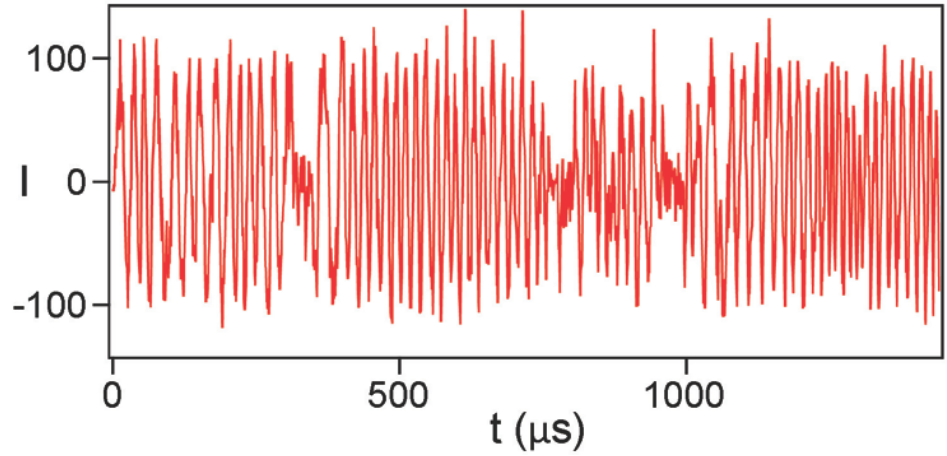

**Fig. S5**
Time-series of the free-running maser emission, $I(t)$. Maser oscillations are interrupted by charge fluctuations, which result in sub-threshold behavior near $t = 300$ µs, $t = 800$ µs, and $t = 950$ µs.

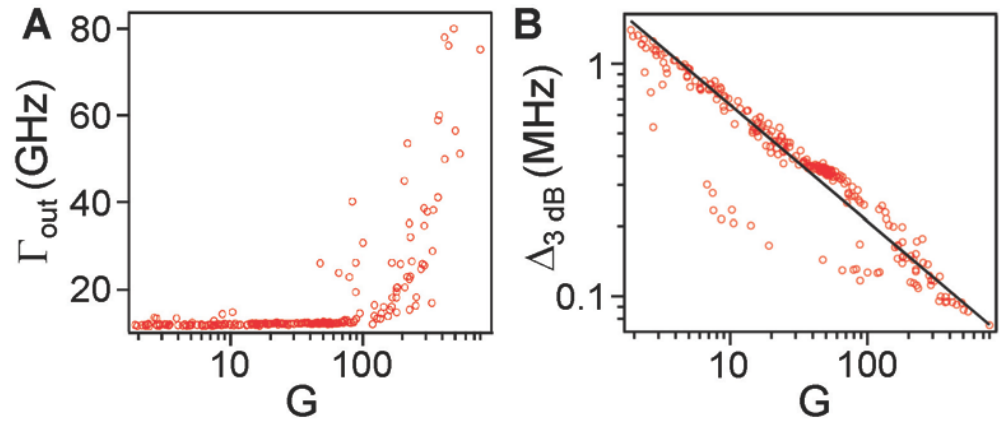

**Fig. S6**
(**A**) Photon emission rate $\Gamma_{out}$ plotted as a function of power gain $G$. (**B**) Emission peak width $\Delta_{3\ dB}$ plotted as a function of $G$ in comparison with the expected gain-bandwidth scaling $\Delta_{3\ dB} \propto \kappa/\sqrt{G}$ (solid line).

**References and Notes**

30. A. Blais, R.-S. Huang, A. Wallraff, S. M. Girvin, R. J. Schoelkopf, *Phys. Rev. A* **69**, 062320 (2004).
31. G.-W. Deng *et al.*, *arXiv:1310.6118*, (2013).
32. H. Toida, T. Nakajima, S. Komiyama, *Phys. Rev. Lett.* **110**, 066802 (2013).
33. C. Eichler *et al.*, *Phys. Rev. Lett.* **106**, 220503 (2011).
34. C. Eichler *et al.*, *Phys. Rev. Lett.* **107**, 113601 (2011).
35. C. Eichler, D. Bozyigit, A. Wallraff, *Phys. Rev. A* **86**, 032106 (2012).
36. J. R. Petta *et al.*, *Science* **309**, 2180 (2005).
37. K. D. Petersson, J. R. Petta, H. Lu, A. C. Gossard, *Phys. Rev. Lett.* **105**, 246804 (2010).
38. L. Childress, A. S. Sorensen, M. D. Lukin, *Phys. Rev. A* **69**, 042302 (2004).
39. M. Kulkarni, O. Cotlet, H. E. Tureci, *Phys. Rev. B* **90**, 125402 (2014).